\documentclass{article}
\usepackage{amsmath,amssymb, amsfonts, theorem}
\newtheorem{lemma}{Lemma}
\newtheorem{prop}{Proposition}
\newtheorem{thm}{Theorem}
\newtheorem{coro}{Corollary}
\newtheorem{defi}{Definition}

\theorembodyfont{\rm}
\newtheorem{ex}{Example}
\newtheorem{rem}{Remark}

\def\Ae {\mathcal A}
\def\Be {\mathcal B}
\def\Ce {\mathcal C}
\def\Ha {\mathcal H}
\def\Ka {\mathcal K}
\def\Le {\mathcal L}

\def\Me {\mathcal M}

\def\Te {\mathcal T}
\def\trans{\mathsf T}

\def\Pe {\mathcal P}
\def\prob{\mathsf P}
\def\Se {\mathfrak S}
\def\mea {\mathbf m}
\def\<{\langle}
\def\>{\rangle}
\def\Tr{{\rm Tr}\,}
\def\ptr{{\rm Tr}}

\def\rank{\mathrm{rank}}

\def\qed{{\hfill $\square$}\vskip 5pt}

\def\<{\langle}
\def\>{\rangle}

\title{Extremal generalized quantum  measurements}
\author{Anna Jen\v cov\'a\\
{\small Mathematical Institute, Slovak Academy of Sciences,}\\
{\small \v {S}tef\'{a}nikova 49, 814 73 Bratislava, Slovakia} \\
{\small  jenca@mat.savba.sk}
}
\date{}

\begin{document}

\maketitle

\noindent
\textbf{Abstract.} A measurement on a section $K$ of the set of states of a finite dimensional $C^*$-algebra
is defined as an affine map from $K$
 to a probability simplex.  Special cases of such sections are 
used in description of quantum networks, in particular quantum channels. Measurements on a section correspond to
 equivalence classes of so-called generalized POVMs, which are called quantum testers in the case of networks.
 We find extremality conditions for 
measurements on $K$ and characterize generalized POVMs such that the corresponding measurement is extremal. 
These results are applied  to the set of channels. We find explicit extremality conditions for two outcome measurements on qubit channels and give an example of an extremal qubit 1-tester such that the corresponding measurement is not extremal.



\section{Introduction}

The motivation for the present work comes from recent papers \cite{daria_circ,daria}, see also \cite{guwat},
where a general framework 
for description of quantum networks
 was developed in terms of positive matrices, also called quantum combs, 
satisfying a set of linear constraints. This description has been useful for
 some important applications, see e.g. \cite{daria_learn,daria_POVM,daria_clonem,daria_cost,daria_clone,gutoski}. In particular, measurements on quantum networks are performed 
by a special kind of networks, which are represented by so-called quantum testers, \cite{daria_testers, ziman}. The extreme points
 of the set of testers were characterized in \cite{daria_extr}.

The set of all combs corresponding
 to a given type of a network forms (a multiple of) a section  of the state space, 
which is an intersection of the set of all positive matrices with unit trace and a linear subspace. 
Motivated by this application, general sections of the state space were studied in \cite{ja}. A 
measurement on a section,
 or a generalized measurement, was
 defined as an affine map from the section to the probability simplex over the set of outcomes. It was proved that 
 measurements are given by  so-called generalized POVMs (positive operator valued measures). In the case of quantum combs, the corresponding generalized POVMs are exactly the  quantum testers.

Since the set of generalized measurements is convex and compact, and since figures of merit for optimalization 
of such measurements are usually convex, it is useful to determine the extreme points. 
Extremal  generalized POVMs were characterized in \cite{ja_extr}, but for general sections, in particular for quantum
 combs, 
 there may exist many generalized POVMs describing the same
 measurement. This defines an equivalence 
relation on generalized POVMs, such that generalized measurements correspond precisely to the equivalence classes.
This means that an extremal generalized POVM  does not necessarily give an  extremal measurement and, on the other hand,
 an extremal generalized measurement can have non-extremal generalized POVMs in its equivalence class.

The aim of the preset paper is  to determine extremal generalized measurements and to  characterize 
generalized POVMs such that the corresponding
 measurement is extremal. For this, we need to describe the largest support projections for  generalized POVMs
 in the same equivalence class.  We also give necessary and sufficient conditions on the support projections 
such that the generalized POVM is unique in its equivalence class.
These results are then applied to the simplest case of a network consisting of a quantum 
 channel with qubit input and output. Moreover, we find 
  an example of an extremal quantum tester such that the corresponding generalized measurement is not extremal.

\section{Notations and preliminaries}

Let $\Ha$ be a finite dimensional Hilbert space and let $\Ae$ be a  $C^*$-subalgebra
in the algebra $B(\Ha)$ of all linear operators on $\Ha$.
The identity in $\Ae$ will be denoted by $I_\Ae$ and 
we fix the trace $\ptr_\Ae$ on $\Ae$ to be the restriction of the trace $\ptr_\Ha$
in $B(\Ha)$, we omit the subscript
$\Ae$ if no confusion is possible. We denote by $\Ae^h$ the set of all self-adjoint elements in $\Ae$
 and by $\Ae^+$ the
convex cone  of positive elements in $\Ae$. If $S\subset \Ae$ is an arbitrary subset, we use the notation
$S^+=S\cap \Ae^+$. If $p\in \Ae$ is a projection, we denote the compressed algebra $p\Ae p$ by $\Ae_p$. 
For $a\in \Ae^+$, the projection onto the support
of $a$ will be denoted by $s(a)$.

 The dual space $\Ae^*$ is usually identified with $\Ae$, with duality
given by $\<a,b\>=\Tr a^*b$. The functional determined by $a\in \Ae$ is  hermitian  if and only if 
$a\in \Ae^h$ and positive if and only if $a\in \Ae^+$. Positive unital functionals are called states and
 are identified with density operators, that is, elements $\rho\in \Ae^+$ with $\Tr\rho=1$. 
We denote the set of states by $\Se(\Ae)$.
If $\Ae=B(\Ha)$, we use the notation $I_\Ha$, $\ptr_\Ha$, $\Se(\Ha)$ etc., with obvious meaning.

Let $\Be$ be another (finite dimensional) $C^*$-algebra, then $\ptr^{\Ae\otimes\Be}_\Ae$, or just $\ptr_\Ae$,
 will
denote the partial trace on the tensor product $\Ae\otimes\Be$, determined by 
$\ptr_\Ae (a\otimes b)=\Tr( a)b$. 	Let $T:\Ae\to \Be$ be a linear map, then $T$ is positive 
if $T(\Ae^+)\subset \Be^+$ and completely positive if the map
\[
T\otimes id_\Le: \Ae\otimes B(\Le)\to \Be\otimes B(\Le)
\]
is positive for all finite dimensional Hilbert spaces $\Le$. A channel $T:\Ae\to \Be$ is 
a completely positive and  trace preserving map.

Any linear map $T:\Ae\to \Be$  is represented by a unique  operator $X_T\in \Be\otimes\Ae$, 
called the Choi matrix of $T$, \cite{choi}. This can be obtained as follows: If $\Ae=B(\Ha)$, then we have
\[
X_T=T\otimes id_\Ha(|\psi_\Ha\>\<\psi_\Ha|),\qquad |\psi_\Ha\>=\sum_i |i\otimes i\>\in \Ha\otimes\Ha
\]
where $|i\>$ denotes an orthonormal basis in $\Ha$. If $\Ae\equiv \oplus_nB(\Ha_n)$, 
then there are maps $T_n:B(\Ha_n)\to\Be$ such that 
$T(a)=T_n(a)$ for $a\in B(\Ha_n)$, and $X_T=\oplus_n X_{T_n}$. The Choi matrix is positive if and only if $T$
 is a completely positive map, and $T$ preserves trace if and only if $\ptr_\Be X_T=I_\Ae$. In this way, the
 set of quantum channels $\Ae\to\Be$ is identified with the set
\[
\Ce(\Ae,\Be):=\{ X\in \Be\otimes\Ae, \ptr_\Be X=I_\Ae\}.
\]

In particular, put $\Ae=B(\Ha)$ and 
$\Be=\mathbb C^m$ and let  $T:\Ae\to \Be$ be a channel. Then $T$ restricts to an affine map from $\Se(\Ha)$ to the probability simplex over the set $\{1,\dots,m\}$. Such maps are
 called measurements on $B(\Ha)$, \cite{holevo}. The Choi matrix of $T$ has the form 
$X_T=\sum_i|i\>\<i|\otimes X_{i}$, with $ X_{i}\in B(\Ha)^+$ and $\sum_i X_{i}=I_\Ha$. We have $T(\rho)(i)=\Tr X_i^t\rho$, where $a^t$ denotes the transpose of $a$.
The relation $T\leftrightarrow X_T$ gives a one-to-one correspondence between measurements and positive operator
 valued measures (POVMs).

\section{Generalized  measurements and generalized\\ POVMs}

 We will fix the following notations throughout the paper: $K$ will denote a  closed convex subset of $\Se(\Ae)$,  
$Q:= \cup_{\lambda\ge 0} \lambda K \subseteq \Ae^+$  the closed convex cone and  $J:=Q-Q\subseteq \Ae^h$  
the real vector subspace generated by $K$. Then $Q$ satisfies $Q\cap -Q=\{0\}$, so that $Q$ defines a partial
 order in $J$.
 
The dual space $J^*$ of $J$ can be identified  
with the quotient of $\Ae^h$, $J^*\equiv \Ae^h|_{K^\perp}$, where 
\[
K^\perp=J^\perp=\{ x\in \Ae^h, \Tr ax=0, \ a\in K\}.
\]
Let $\pi:\Ae^h\to J^*$ be 
the quotient map, $\pi(a)=a+K^\perp$. Then the duality of $J$ and $J^*$ is given by 
\[
\< \pi(a),x\>=\Tr ax,\qquad a\in \Ae^h,\ x\in J.
\]
A linear functional on $J$ is positive if its value is positive on every element of $Q$. The set
 of all positive functionals is the dual cone $Q^*$. This is  a closed convex cone in $J^*$  and 
 $Q^{**}=Q$. Since $Q\subseteq \Ae^+$, we always have $\pi(\Ae^+)\subseteq Q^*$. 
\begin{thm}\label{thm:sect} \textrm{\cite{ja}} $Q^*=\pi(\Ae^+)$ if and only if $K=J\cap \Se(\Ae)$.

\end{thm}
\begin{defi} A subset $K\subseteq \Se(\Ae)$ satisfying $K=J\cap \Se(\Ae)$  will be called a section of $\Se(\Ae)$.
\end{defi}
 We will next show some important examples of sections.

\begin{ex}(\textbf{Quantum channels})\label{ex:chan}  
Let $K=(\Tr I_\Ae)^{-1}\Ce(\Ae,\Be)$, then it is not difficult to see that
$K$ is a section of $ \Se(\Be\otimes\Ae)$, \cite{ja}.
In this case, $J=\{X\in (\Be\otimes \Ae)^h, \ptr_\Be X=tI_\Ae, t\in \mathbb R\}$ and 
$K^\perp=I_\Be\otimes \{I_\Ae\}^\perp$, note that $\{I_\Ae\}^\perp$ is the set of elements in $\Ae^h$ with zero trace.
\end{ex}

\begin{ex}(\textbf{Quantum supermaps})\label{ex:superm} Quantum supermaps were introduced in 
\cite{daria_super} as completely positive maps that map the set $\Ce(\Ha_0,\Ha_1)$ into $\Ce(\Ka_0,\Ka_1)$.
The following definition was used in \cite{ja}.

Let $\Be_0,\Be_1,\dots$ be a sequence of finite dimensional $C^*$ algebras. We denote by $\Ce(\Be_0,\Be_1,\dots,\Be_n)$
 the set of Choi matrices of  completely positive maps $\Be_{n-1}\otimes\dots\otimes\Be_0\to \Be_n$ such that
$\Ce(\Be_0,\dots,\Be_{n-1})$ is mapped into $\Se(\Be_n)$.  If $n$ is odd, then $\Ce(\Be_0,\dots,\Be_n)$ corresponds to the set of (conditional)
 quantum combs, which are defined as Choi matrices of completely positive maps 
$\Be_{n-1}\otimes\dots\otimes\Be_1\to \Be_n\otimes \Be_0$,
such that $\Ce(\Be_1,\dots,\Be_{n-1})$ is mapped into $\Ce(\Be_0,\Be_n)$. Quantum combs are used for representation 
 of  general quantum networks, \cite{daria_circ,daria,guwat}. The most general form of a conditional comb was introduced 
 in \cite{daria_cond}.

As it was shown in \cite{daria},  quantum supermaps are quantum channels whose  Choi matrices
satisfy a set of linear constraints. This implies that $\Ce(\Be_0,\dots,\Be_n)$
 is again (a multiple of) a  section of a state space.

\end{ex}

Next we define a generalized quantum measurement with  values in a finite set $U$. Let $\prob(U)$ denote
 the probability simplex over $U$.

\begin{defi} Let $K\subseteq \Se(\Ae)$ be a closed convex set. A generalized measurement on $K$ with values
 in a (finite) set $U$ is an affine map  $K\to\prob(U)$. The set of all generalized  measurements on $K$ will be
denoted by $\Me(K,U)$.

\end{defi}

 It is  easy to see that any measurement
$\mea:K\to \prob(U)$ is given by elements $\mea_u\in Q^*$, $\mea_u(\rho):=\mea(\rho)(u)$, $\rho\in K$, $u\in U$,
 and we must have  $\sum_{u\in U} \mea_u=\pi(I)$.
Let $M_u\in \Ae^h$  be such that $\pi(M_u)=\mea_u$, then we have $\sum_uM_u\in \pi(I)=I+K^\perp$.
Conversely, it is clear that  any collection of \emph{positive} operators satisfying this condition defines a generalized
 measurement.

\begin{defi} \textrm{\cite{ja}} A generalized POVM (with respect to $K$) is a collection of positive operators $\{M_u, u\in U\}$
 such that $\sum_uM_u\in I+K^\perp$. The set of all generalized POVMs will be denoted by $\Me_K(\Ae,U)$.

\end{defi}

From now on, we will always assume that $K$ is a section of $\Se(\Ae)$.
By Theorem \ref{thm:sect}, generalized   measurements in this case correspond precisely to  equivalence classes of 
generalized POVMs. 
If $M=\{M_u, u\in U\}\in \Me_K(\Ae,U)$, the corresponding measurement is $\pi(M):=\{\pi(M_u), u\in U\}$.

\begin{ex}\label{ex:testers}\textbf{(Quantum testers)}
Since $K=\Tr(I_\Ae)^{-1}\Ce(\Ae,\Be)$ is a section, any generalized measurement on the set of quantum channels 
is given by a generalized POVM with respect to $K$, multiplied by $\Tr(I_\Ae)^{-1}$. We obtain a collection $\{M_u\}$ of positive 
operators such that $\sum_u M_u=I_\Be\otimes \sigma$ 
for some $\sigma\in \Se(\Ae)$. Such objects were studied in \cite{daria_testers, daria, ziman} and  called quantum 1-testers, 
or PPOVMs in \cite{ziman}. We will denote the set of all
 quantum 1-testers by $\Te(\Ae,\Be,U)$. More generally,  generalized POVMs for quantum supermaps are  called 
quantum testers (see  \cite{daria_testers, daria}), note that by definition,
 quantum testers are elements in $\Ce(\Be_0,\dots,\Be_n,\mathbb C^{|U|})$ and hence are quantum supermaps 
themselves.

\end{ex}

Let $M\in \Me_K(\Ae,U)$. The following decomposition was used  for quantum testers in \cite{daria_testers}
 and in \cite{ja, ja_extr} for generalized POVMs: 
Let $\sum_u M_u=c$ and let  $p=s(c)$. Let us define $\chi_c:\Ae
\to \Ae$ by $a\mapsto c^{1/2}ac^{1/2}$. Then $\chi_c$ is completely positive and preserves trace on $J$.
Moreover, restricted to the compressed algebra $\Ae_p$, $\chi_c$ has an inverse $\chi_c^{-1}=\chi_{c^{-1}}$.
Put $\Lambda_u=\chi_c^{-1}(M_u)$ (note that $s(M_u)\le p$, so that $M_u\in \Ae_p$). Then 
$\Lambda=\{\Lambda_u, u\in U\}$ is a POVM on  $\Ae_p$, 
 such that 
\[
\Tr M_u\rho=\Tr \Lambda_u\chi_c(\rho),\qquad u\in U,\ \rho\in K.
\]
This decomposition will be written as $M=\Lambda\circ\chi_c$.

For quantum 1-testers, this has the following form \cite{daria_testers,ziman}:
Suppose  that $\Ae=B(\Ha)$ and let $M=\{M_u\}$, $M_u\in \Be\otimes B(\Ha)$ be a quantum 1-tester, with
 $\sum_u M_u=I_\Be\otimes \sigma$, $\sigma\in \Se(\Ha)$.
 Let $\chi=\chi_{I_\Be\otimes \sigma}$ and let $q=s(\sigma)$ and $\Le=q\Ha$, so that $M=\Lambda\circ \chi$ 
with a POVM $\Lambda=\{\Lambda_u, u\in  U\}$ on $\Be\otimes B(\Le)$.
Then  for a channel $T:\Ae\to \Be$ with Choi matrix $X_T$, 
\[
\chi(X_T)=(T\otimes id_{\Le})(\xi)
\]
where $\xi:=(I_\Ha\otimes \sigma^{1/2})|\psi_\Ha\>\<\psi_\Ha|(I_\Ha\otimes \sigma^{1/2})$ is a pure state with $\ptr_{\Le}\xi=\sigma^\trans$. 
Hence the tester $M$ has an implementation $\Tr M_uX_T=\Tr \Lambda_u(T\otimes id_\Le)(\xi)$.

\section{Extremality conditions}

Since $K$ contains an element with largest support, we may always assume that $K$ contains an element of full rank, by restriction
 to a compressed algebra. In this case, it follows from \cite[Proposition 6]{ja} that $\Me_K(\Ae,U)$ is compact, and obviously also convex. The extreme points of the set of testers were obtained in \cite{daria_extr} and for generalized POVMs in \cite{ja_extr}.  
The proposition below sumarizes some of the results.

Let $M=\Lambda\circ \chi_c$ with $p:=s(c)$. Let  $K_c:=\chi_c(J)\cap \Se(\Ae)$.
It is clear that the POVM
$\Lambda$ is a generalized POVM with respect to any section, hence in particular, 
$\Lambda\in \Me_{K_c}(\Ae_p,U)$.

\begin{prop}\label{prop:extr_tester} \textrm{\cite{ja_extr} }
\begin{enumerate}
\item[(i)] Let $p_u=s(M_u)$. Then $M$ is extremal in $\Me_K(\Ae,U)$ if and only if 
for any collection of operators $D_u\in \Ae^h_{p_u}$,  $\sum_u D_u\in J^\perp$ implies that $D_u=0$
 for all $u\in U$.

\item[(ii)]  $M$ is extremal 
 in $\Me_K(\Ae,U)$ if and only if $\Lambda$ is extremal in $\Me_{K_c}(\Ae_p,U)$. 
\item[(iii)] Let $M$ be a quantum 1-tester, $M\in \Te(\Ha,\Ka,U)$, and let $c=I_\Ka\otimes \sigma$ for 
 $\sigma\in \Se(\Ha)$, $s(\sigma)=q$. Then $M$ is  extremal in $\Te(\Ha,\Ka,U)$ 
if and only if $\Tr(q)^{-1} \Lambda$ is 
 extremal   in $\Te(q\Ha,\Ka,U)$. 
\item[(iv)] If $|U|=2$ in (iii),  so that $M$ is a 1-tester with 2 outcomes, then $M$ is  extremal if and only if
 $\Lambda=(\Lambda_1,\Lambda_2)$ with $\Lambda_1$ a projection not commuting with any projection of the form 
$I_\Ka\otimes e$ with $e\neq 0,q$.

\end{enumerate}
\end{prop}

\begin{rem} Note that if $K=\Se(\Ae)$, hence in the case of ordinary POVMs, the condition (i) becomes  weak independence of the supports of 
$M_u$, $u\in U$. This extremality condition for POVMs was obtained by Arveson \cite{arv} in a very general infinite dimensional 
seting. In finite dimensions, this condition was proved by a perturbation method in \cite{daria_bary}.

\end{rem}

\begin{ex}\label{ex:extremal_tester}
For $\dim(\Ha)=2$, all extreme points in $\Te(\Ha,\Ka,\{0,1\})$ can be characterized as follows 
 \cite{daria_extr,ja_extr}: Let $M=\Lambda\circ \chi_{I\otimes\sigma}$.
 If $\rank(\sigma)=1$, then $M$ is extremal if and only if $M$ is a PVM. If $\rank(\sigma)=2$, then 
$M$ is extremal if and only if $\Lambda$ is a PVM and $\Lambda_0$ (and hence also $\Lambda_1$)
 is  not of the form
\begin{equation}\label{eq:lambda_notex}
\Lambda_0=e\otimes |\psi\>\<\psi|+f\otimes |\psi^\perp\>\<\psi^\perp|,
\end{equation}
where $\psi,\psi^\perp\in \Ha$ are orthogonal unit vectors and $e,f$ are projections on $\Ka$.

\end{ex}

We now turn to extremality conditions  for generalized measurements.
Since the set $\Me(K,U)$ is the image of $\Me_K(\Ae,U)$ under the linear
map $\pi$, it must be convex and compact as well. We will now characterize the extreme points.
First, let $\mathbf a$ be any element in $Q^*$, so that $\mathbf a=\pi(a)$ for some $a\in \Ae^+$. 
Consider the set $(a+K^\perp)^+$ 
of all positive elements
 in the equivalence class  of $a$. Since this is  a closed convex subset in $\Ae^+$,  it contains
 some element $b$ with largest support. Let us denote $s(\mathbf a):=s(b)$.

\begin{thm}\label{thm:extr_sec} Let $\mea\in \Me(K,U)$ and let $s(\mea_u)=s_u$, $u\in U$. 
Then  $\mea$ is extremal if
 and only if for any collection $\{\mathbf x_u\in \pi(\Ae^h_{s_u}), u\in U\}$, 
 $\sum_u\mathbf x_u=\pi(0)$ implies that $\mathbf x_u=\pi(0)$ for all $u\in U$.

\end{thm}

{\it Proof.} The proof uses the standard perturbation method of convex analysis.
So let us suppose that $\mea$ is extremal in $\Me(K,U)$ and let $\{\mathbf x_u\in \pi(\Ae_{s_u}^h), u\in U\}$
 be such that $\sum_u \mathbf x_u=\pi(0)$. Let $M_u$ be such that $\pi(M_u)=\mea_u$ and $s(M_u)=s_u$; and
choose any  $X_u\in \Ae_{s_u}^h$  such that $\pi(X_u)=\mathbf x_u$. Then there is some $s>0$ satisfying
$M_{\pm,u}:= M_u\pm sX_u\in \Ae^+$, for all $u\in U$. Hence $\mea_{\pm,u}:= \pi(M_{\pm,u})\in Q^*$, moreover, $\sum_u \mea_{\pm,u}=
\sum_u \mea_u=\pi(I)$. Since $\mea=\frac12(\mea_++\mea_-)$ and $\mea$ is extremal, this implies that
$\mea_+=\mea_-$, hence $2sX_u=M_{+,u}-M_{-,u}\in J^\perp$, so that $\mathbf x_u=\pi(0)$ for all $u\in U$.

Conversely, suppose the condition is satisfied and let $\mea=\frac12(\mea^1+\mea^2)$.
Let $M^i$ be generalized POVMs such that  $\pi(M^i)=\mea^{i}$, $i=1,2$, then $\pi(\frac 12( M^1+M^2))=\mea$, therefore 
$s(M^i_{u})\le s(M^1_u+M^2_u)\le s(\mea_u)$, so that $M^i_{u}\in \Ae^h_{s_u}$. Hence $\mathbf x_u=\pi(M^1_u-M^2_{u})\in \pi(\Ae^h_{s_u})$, moreover, $\sum_u \mathbf x_u=\pi(0)$. It follows that $\mathbf x_u=\pi(0)$ for all $u$ 
and hence $\mea^1=\pi(M^1)=\pi(M^2)=\mea^2$.

\qed

\begin{coro} Let $\mathbf m\in \Me(K,U)$ be extremal and let $s(\mathbf m_u)=s_u$.   Then 
\[
\sum_u \dim (s_uJs_u)\le \dim (J).
\]

\end{coro}

{\it Proof.} Let us denote $J^*_U=\oplus_{u\in U} J^*$ and let 
$L=\{\mathbf x\in J^*_U, \sum_u\mathbf x_u=\pi(0)\}$. Then the extremality condition in
Theorem \ref{thm:extr_sec} has the form
\[
\oplus_{u\in U}\pi(\Ae_{s_u})\cap L=\{\pi(0)_U\},
\]
where $\pi(0)_U$ is the zero element in $J^*_U$. By taking  orthogonal complements, we obtain 
that
\[
(\oplus_{u\in U} \pi(\Ae^h_{s_u}))^\perp\vee  L^\perp=J^*_U,
\]
which implies that 
\[
\sum_u \dim(\pi(\Ae^h_{s_u}))=\dim(\oplus_u \pi(\Ae^h_{s_u}))\le \dim(L^\perp).
\]
Note that $s_uJs_u$ is a subspace in $\Ae^h_{s_u}$ and it is easy to see that 
$(s_uJs_u)^{\perp}\cap \Ae^h_{s_u}=J^\perp\cap \Ae_{s_u}$.
As before, we may identify the dual space with the quotient
 space $(s_uJs_u)^*\equiv
\Ae^h_{s_u}|_{J^\perp\cap\Ae^h_{s_u}}$. Let 
$\pi_{u}$ be the quotient map, then  for elements $x,y\in \Ae^h_{s_u}$, 
$\pi(x)=\pi(y)$ if and only if $\pi_{u}(x)=\pi_{u}(y)$. It follows that 
\[
\dim(\pi(\Ae^h_{s_u}))=\dim(\pi_{u}(\Ae^h_{s_u}))=
\dim((s_uJs_u)^*)=\dim(s_uJs_u).
\]
Moreover, it is easy to check that $L^\perp=\{y\in J_U:=\oplus_u J, y_u=y_v, u,v\in U\}\equiv J$, so that $\dim(L^\perp)=\dim(J)$.

Putting this together, we obtain the statement.

\qed

We now characterize generalized POVMs corresponding to an extremal measurement. For $a\in \Ae^+$, let
$s_K(a):=s(\pi(a))$, then $s_K(a)$ is the largest support of an element in $(a+K^\perp)^+$.
We  call $s_K(a)$ the $K$-support of $a$. The next statement follows directly
 from Theorem \ref{thm:extr_sec} (compare  with Proposition \ref{prop:extr_tester} (i)).

\begin{thm}\label{thm:gPOVM_extr_meas} Let $M\in \Me_K(\Ae,U)$ and let $s_u=s_K(M_u)$, $u\in U$.
Then $\pi(M)$ is extremal in $\Me(K,U)$ if and only if
 for any $\{D_u\in\Ae^h_{s_u},u\in U\}$, $\sum_u D_u\in K^\perp$ implies that
 $D_u\in K^\perp$ for all $u\in U$.

\end{thm}

To make the above characterization more useful, we need to describe the $K$-supports of positive elements
in $\Ae$.

\begin{prop}\label{prop:Ksup} Let $a\in\Ae^+$. Then $s(a)=s_K(a)$ if and only if there is an element $b\in Q$
 such that $s(a)=I-s(b)$.

\end{prop}

{\it Proof.} Let $p=s(a)$. Suppose $p=s_K(a)$. Note that we have
\begin{equation}\label{eq:pom}
\{x\in \Ae^h,\ \exists t>0,\ a+tx\in \Ae^+\}=\Ae^++\Ae^h_p
\end{equation}

Let now $x\in (\Ae^++\Ae^h_p)\cap K^\perp$, then (\ref{eq:pom}) implies that there is some $t>0$ such that
$a_0:=a+tx\in (a+K^\perp)^+\subseteq \Ae_p^+$, since $p=s_K(a)$. Hence $x=t^{-1}(a_0-a)
\in \Ae^h_p$, so that 
$(\Ae^++\Ae^h_p)\cap K^\perp\subseteq \Ae^h_p\cap K^\perp$. Since the converse inclusion is clear, we have
\[
(\Ae^++\Ae^h_p)\cap K^\perp= \Ae^h_p\cap K^\perp
\]
Applying the duality $^*$ of the convex cones to this equality, 
we get (\cite[Corollary 11.4.2]{rockafellar})
\[
cl((\Ae^++\Ae^h_p)^*+J)= cl(\Ae^h_{I-p}+J).
\]
Since $\Ae^h_{I-p}+J$ is an 
affine subspace, we may remove the closure operator and we get from
$(\Ae^++\Ae^h_p)^*=\Ae^+\cap \Ae_{I-p}^h=\Ae^+_{I-p}$ that
\[
\Ae^+_{I-p}+J=\Ae_{I-p}^h+J
\]
In particular, we have 
$-\Ae^+_{I-p}\subseteq \Ae^+_{I-p}+J$. Let $c\in \Ae^+$, $s(c)=I-p$, then there 
are some $d\in \Ae^+_{I-p}$ and $x\in J$ such that
$-c=d+x$. But then $b:=c+d\in Q$, $s(b)=I-p$.

Conversely, let $b\in Q$, $s(b)=I-p$,  then 
\[
\Ae_p^+=\{b\}^\perp\cap \Ae^+\supseteq (\Ae_p^++K^\perp)^+\supseteq \Ae_p^+
\]
so that $\Ae_p^+=(\Ae_p^++K^\perp)^+$. In particular, $(a+K^\perp)^+\subseteq \Ae_p^+$,
 so that $p=s_K(a)$.

\qed

Let us denote 
\[
\Pe_K(\Ae):=\{I-s(b), b\in Q\}.
\]
Note that for any subset $\Pe\subset \Pe_K(\Ae)$, we have $\bigwedge \Pe\in \Pe_K(\Ae)$, so that 
$\Pe_K(\Ae)$ is a $\wedge$-complete semilattice. Indeed, let
$R\subset Q$ be the set of elements such that $\Pe=\{I-s(a), a\in R\}$, then 
there is some $b$ in the closed convex hull of $R$, such that $s(b)=\bigvee \{s(a), a\in R\}$ so that 
$I-s(b)=\bigwedge \{I-s(a), a\in R\}=\bigwedge \Pe$. Since 
 $b\in Q$, we have  $I-s(b)\in \Pe_K$.

\begin{prop}
 Let $a\in \Ae^+$. Then
  $s_K(a)=\bigwedge \{ s\in \Pe_K(\Ae),\  s(a)\le s\}$.
\end{prop}

{\it Proof.}  By Proposition \ref{prop:Ksup}, $s_K(a)\in \Pe_K(\Ae)$ and $s(a)\le s_K(a)$ by definition.
Let $s'$ be another such projection,
with $1-s'=s(b')$, $b'\in Q$. Then  
  $\Tr b' d=\Tr b'a=0$ for all $d\in (a+K^\perp)^+$. This implies $s(d)\le 1-s(b')=s'$, so that 
$s_K(a)\le s'$.

\qed

\begin{rem}\label{rem:faces} Let $p\in \Ae$ be a projection, then it is easy to see that 
the set $\{\mathbf a\in Q^*, s(\mathbf a)\le p\}$ is a face of $Q^*$. Conversely, any face of $Q^*$ has this form:
  if $F\subset Q^*$ is a face, then $\pi^{-1}(F)\cap \Ae^+$ is a face of $\Ae^+$, hence  $\pi^{-1}(F)\cap \Ae^+=\Ae_p^+$ 
for some projection $p$. Consequently, $F=\pi(\Ae^+_p)=\{\mathbf a\in Q^*, s(\mathbf a)\le p\}$. By Proposition \ref{prop:Ksup}, there is a 1-1 correspondence between faces of $Q^*$ and $\Pe_K(\Ae)$.

Similarly, faces of $\Me(K,U)$  are the sets  $\{\mathbf m\in \Me(K,U), s(\mathbf m_u)\le p_u, u\in U\}$ for some
 projections $p_u$ and there is a 1-1 correspondence between faces of $\Me(K,U)$ and $U$-tuples $\{p_u,u\in U\}$, $p_u\in \Pe_K$. In particular,
$\Me_p(K,U):=\{\mathbf m\in \Me(K,U), s(\mathbf m_u)\le p, u\in U\}$ is a face of $\Me(K,U)$, for any projection $p$.

\end{rem}

Suppose that $M\in \Me_K(\Ae,U)$ has the decomposition   $M=\Lambda\circ \chi_c$, with $p=s(c)$.
We now relate extremality of the measurement given by $M$ to  extremality of the measurement given by $\Lambda$,
 cf. Proposition \ref{prop:extr_meas_qchans} (ii).  Let $K_c:=\chi_c(J)\cap\Se(\Ae_p)$, $Q_c:=\chi_c(J)\cap \Ae_p^+$ and let
 $\pi_c$ be the corresponding quotient map 
$\pi_c: \Ae_p^h\to \Ae_p^h|_{\chi_c(J)^\perp}$.  Note that we have $p=\vee_u s(M_u)\le \vee_u s_K(M_u)$.
\begin{thm} \label{thm:extremal_decomp}
Suppose that  $p=\vee_u s_K(M_u)$.
Then $\pi(M)$ is extremal in  $\Me(K,U)$ if and only if $\pi_{c}(\Lambda)$ is extremal in $\Me(K_c,U)$.

\end{thm}

{\it Proof.} Let  $a\in \Ae^+$ be any element such that $s_K(a)\le p$. Then $(a+K^\perp)^+\subset \Ae_p^h$, so that 
$(a+K^\perp)^+=(a+(K^\perp\cap \Ae_p^h))^+$ and it
 is easy to check that
 $\chi_c^{-1}(K^\perp\cap \Ae^h_p)=K_c^\perp\cap \Ae_p^h$. It follows that $\chi_c^{-1}$
maps $(a+K^\perp)^+$ onto $(\chi_c^{-1}(a)+K_c^{\perp})^+$ and hence  $\pi(a)\mapsto \pi_c(\chi_c^{-1}(a))$
defines an invertible affine  
map from the face $\{\mathbf a\in Q^*, s(\mathbf a)\le p\}$ onto $Q_c^*$.  
 
 Moreover, let $d\in (I+K^\perp)\cap \Ae_p^+$ and let $\rho_c\in K_c$, so that $\rho_c$ is a state of the form 
 $\rho_c=\chi_c(x)$ for some $x\in J$ and we have $1=\Tr \rho_c =\Tr xc =\Tr x$ (since $c\in I+K^\perp$). Then
 \[
\Tr \chi_c^{-1}(d)\rho_c=\Tr dpxp=\Tr dx=\Tr x=1,
 \]
 hence  $\chi_c^{-1}(d)\in (p+K_c^\perp)^+$. By a similar argument, we can see that $\chi_c(p+K_c^\perp)^+=(I+K^\perp)\cap \Ae_p^+$.
From this, one can see that $\chi_c^{-1}$ defines an affine   invertible  map from 
the face $\Me_p(K,U)$ onto $\Me(K_c,U)$, see Remark \ref{rem:faces}. Since $\pi(M)\in \Me_p(K,U)$
 and $\chi_c^{-1}(\pi(M))=\pi_c(\Lambda)$, 
the statement follows.
 
 \qed

\subsection{Equivalence of generalized POVMs}

In this paragraph, we deal with the question whether  a given generalized POVM $M$ is the unique element
 in its equivalence class. For this, it is enough to characterize the situation when 
 $(a+K^\perp)^+=\{a\}$ for $a\in \Ae^+$.

\begin{lemma} Let $a\in \Ae^+$, then  $(a+K^\perp)^+$ is convex and compact.

\end{lemma}

{\it Proof.}
It is enough to show that $(a+K^\perp)^+$
is bounded in some norm. Let $\rho\in K$ be of full rank, then for all $b\in (a+K^\perp)^+$,
$\|\rho^{1/2}b\rho^{1/2}\|_1=\Tr \rho b=\Tr \rho a=:t$. Hence $\|b\|_1\le \|\rho^{-1}\|t$ and $(a+K^\perp)^+$ is bounded.

\qed
  \begin{lemma}\label{lemma:extremal} Let $a\in \Ae^+$, $s=s(a)$. 
Then $a$ is extremal in $(a+K^\perp)^+$ if and only if
 $\dim(sJs)=\dim(\Ae_s^h)$.

  \end{lemma}
{\it Proof.} By applying the perturbation method, it is easily seen that $a$ is extremal in 
$(a+K^\perp)^+$ if and only if $K^\perp\cap \Ae_s^h=\{0\}$. Since $K^\perp\cap \Ae_s^h=(sJs)^\perp\cap
\Ae_s^h$, this  is 
equivalent with  $\dim(sJs)=
\dim(\Ae_s^h)$.

\qed

\begin{prop}\label{prop:unique_in_its_class} Let $a\in \Ae^+$, $s=s(a)$. Then  $a$ is the unique positive element in its equivalence class if and only if
$s\in \Pe_K(\Ae)$ and $\dim(sJs)=\dim(\Ae_s^h)$.
\end{prop}

{\it Proof.} The conditions  say that $a$ is an extreme point in $(a+K^\perp)^+$
 such that $s(a)$ contains the supports of all other elements in  $(a+K^\perp)^+$. This 
happens if and  only if
$a$ is the unique point in this set.

\qed

\section{Extremal measurements on qubit channels}

We now apply the results of the previous section to the set $\Te(\Ha,\Ka,U)$ of quantum 1-testers
 for finite dimensional Hilbert spaces $\Ha$ and $\Ka$.
Let $c=I_\Ka\otimes \sigma$, $\sigma\in \Se(\Ha)$ and let $J$ be as in Example \ref{ex:chan}. Then
\[
K_c=\chi_c(J)\cap \Se(\Ka\otimes\Ha)=\{\rho\in \Se(\Ka\otimes\Ha), \ptr_\Ka \rho=\sigma\}.
\]
Note that for $\sigma=\dim(\Ha)^{-1}I_\Ha$, $K_c=K=\dim(\Ha)^{-1}\Ce(\Ha,\Ka)$.  
  Let $p\ne I$ be a projection on $\Ka\otimes\Ha$. Then one can see from 
the definition that 
$p\in \Pe_{K_c}(\Ka\otimes\Ha)$ if and only if there are one-dimensional projections $p_i=|\phi_i\>\<\phi_i|$, $i=1,\dots,k$ such that $1-p=\vee_i p_i$ and 
the convex hull  $co\{\ptr_\Ka p_1,\dots,\ptr_\Ka p_k\}$ contains $\sigma$.
In particular, if $1-p=|\phi\>\<\phi|$, then $p\in \prob_K(\Ka\otimes\Ha)$ if and only if $|\phi\>\<\phi|$ is maximally entangled.

While it is not easy to describe the sets $(a+K_c^\perp)^+$, we can establish the following simple facts.

\begin{lemma}\label{lemma:gjrho} Let $a\in B(\Ka\otimes\Ha)^+$ and let $c=I_\Ka\otimes\sigma$ with $\sigma\in \Se(\Ha)$  of full rank.
\begin{enumerate}
\item[(i)] If $\rank(a)<\dim(\Ka)$ then
$(a+K_c^\perp)^+=\{a\}$.
\item[(ii)]  If $\rank(a)<2\dim(\Ka)$ then 
$a$ is an extreme point in $(a+K_c^\perp)^+$.
\item[(iii)] If 
$\rank(a)^2>\dim(\Ha)^2\dim(\Ka)^2-\dim(\Ha)^2+1$,
 then $(a+K_c^\perp)^+\ne \{a\}$.

\end{enumerate}
\end{lemma}

{\it Proof.} It is clear that  $K_c^\perp=I_\Ka\otimes \{\sigma\}^\perp$.

 (i) Suppose $b\in (a+K_c^\perp)$, $b\ge 0$ and  $b\ne a$, then there is some 
nonzero $y\in \{\sigma\}^\perp$, such that 
$b=a+I_\Ka\otimes y$. Let $y=y_+-y_-$ be the decomposition of $y$ into its positive and negative part, that is, 
$y_\pm\in B(\Ha)^+$ and $s(y_+)s(y_-)=0$. Then we have
\[
I_\Ka\otimes y_-\le a+I_\Ka\otimes y_+
\]
Since we have positive elements on both sides, this implies that
\[
I_\Ka\otimes s(y_-)\le s(a+I_\Ka\otimes y_+)=s(a)\vee (I_\Ka\otimes s(y_+))
\]
and since $s(y_+)$ and $s(y_-)$ are orthogonal projections, we must have 
$\rank(I_\Ka\otimes s(y_-))\le \rank (s(a))$. It follows that
\[
\rank(a)=\rank(s(a))\ge \rank(I_\Ka\otimes s(y_-))\ge \dim(\Ka).
\] 
Hence $\rank(a)<\dim(\Ka)$ implies $(a+K_c^\perp)^+=\{a\}$.

(ii) Let $s=s(a)$ and let  $0\ne z\in B(s(\Ka\otimes\Ha))^h\cap K_c^\perp$, then  $z=I_{\Ka}\otimes y$ for 
$y\in \{\sigma\}^\perp$.  Since $y\ne 0$, $\rank(y)$ must be at least 2. Hence 
$\rank(z)\ge 2\dim(\Ka)$ and we must have $\rank(a)=\rank(s)\ge \rank(z)$. 
Hence  $\rank(a)<2\dim(\Ka)$ implies that $B(s(\Ka\otimes\Ha))^h\cap K_c^\perp=\{0\}$.
By the proof of Lemma \ref{lemma:extremal},
$a$ is then  an extreme point in $(a+ K_c^\perp)^+$.

(iii) Let $J_c=\chi_c(J)$, then $J_c$ is the real linear span of $K_c$.
By Proposition \ref{prop:unique_in_its_class}, $(a+K_c^\perp)^+=\{a\}$ implies 
\[
\rank(a)^2=\dim(B(s(\Ka\otimes \Ha))^h)=\dim(sJ_cs)\le \dim(J)
\]
and $\dim (J)=\dim(B(\Ka\otimes \Ha)^h)-\dim(J^\perp)=\dim(\Ha)^2\dim(\Ka)^2-\dim(\Ha)^2+1$.

\qed

\begin{lemma}\label{lemma:gjrho_qubit} Suppose $\dim(\Ha)=\dim(\Ka)=2$ in the previous lemma. Then
 $(a+K_c^\perp)^+\ne \{a\}$ if and only if 
$s_{K_c}(a)=I$.

\end{lemma}

{\it Proof.}  Let 
$b\in (a+K_c^\perp)^+$ be such that $s(b)=s_{K_c}(a)$. If $\rank(b)<4$ then by Lemma 
\ref{lemma:gjrho}
 (ii), $b$ is an extreme point in $(b+K_c^\perp)^+$, so that 
$(a+K_c^\perp)^+=(b+K_c^\perp)^+=\{b\}$ has exactly one element. The converse follows by Lemma 
\ref{lemma:gjrho} (iii).

\qed

We can now characterize extremal generalized measurements for the set of qubit channels.

\begin{prop}\label{prop:extr_meas_qchans}  Let $\dim(\Ha)=\dim(\Ka)=2$, $M\in \Te(\Ha,\Ka,U)$. Then $\pi(M)$ is  extremal
   if and only if $M$ is extremal in $\Te(\Ha,\Ka,U)$ and $s(M_u)\in \prob_K(\Ka\otimes\Ha)$ for all $u\in U$. 
\end{prop}

{\it Proof.} Suppose that $\pi(M)$ is extremal. Then Theorem \ref{thm:gPOVM_extr_meas} implies that $s_K(M_u)$ cannot be equal 
to $I_{\Ka\otimes\Ha}$ for any $u\in U$.
By Lemma \ref{lemma:gjrho_qubit}, this implies $(M_u+K^\perp)^+=\{M_u\}$. It follows that $M$ is an extremal 1-tester and 
$s(M_u)=s_K(M_u)\in \Pe_K(\Ka\otimes\Ha)$ for all $u$.

Conversely, extremality of $M$ and the fact that $s(M_u)=s_K(M_u)$ imply that $M$ is unique in its equivalence 
class and hence the corresponding measurement must be extremal as well.

\qed

We will next characterize extremality of $\pi(M)$ in terms of the implementing POVM.
So let $M=\Lambda\circ\chi$, $\chi=\chi_{I\otimes\sigma}$, be the decomposition of $M$ and let $q=s(\sigma)$.

\begin{coro}\label{coro:extr_meas_qchans_lambda}  $\pi(M)$ is an extremal measurement on qubit channels 
 if and only if $\Tr(q)^{-1} \Lambda$ is extremal in $\Te(q\Ha,\Ka,U)$ and $s(\Lambda_u)\in \Pe_{K_c}
(\Ka\otimes q\Ha)$ for all $u\in U$.
\end{coro}

 {\it Proof.} Suppose first that $\sigma=|\varphi\>\<\varphi|$ for some $\varphi\in \Ha$. Then $M_u=N_u\otimes|\varphi\>\<\varphi|=\Lambda_u$ for some POVM $N_u$ on $B(\Ka)$ and $K_c=\Se(\Ka\otimes|\varphi\>)$, so that the assertion follows by Proposition 
 \ref{prop:extr_meas_qchans}. 
 
If $\rank(\sigma)=2$, then the assertion follows by Theorem \ref{thm:extremal_decomp}, Proposition \ref{prop:extr_tester} (iii)  and Lemma \ref{lemma:gjrho_qubit} similarly 
as in the proof of Proposition \ref{prop:extr_meas_qchans}.

\qed

The next Example shows an extremal qubit 1-tester, such that the corresponding 
measurement is not extremal.

\begin{ex}\label{ex:extremal_qubit}
Let us apply the above results to the case of two outcomes. Let $M=(M_1,M_2)$ be a qubit 1-tester with 
$M_1+M_2=I\otimes\sigma$,
 where $\rank(\sigma)=2$ and let $M=\Lambda\circ\chi$. Then by Corollary \ref{coro:extr_meas_qchans_lambda} and Example \ref{ex:extremal_tester}, 
 $\pi(M)$ is extremal if and only if 
 $\Lambda_1,\Lambda_2\in \Pe_{K_c}(\Ka\otimes\Ha)$ and $\Lambda_1$ is not of the form
(\ref{eq:lambda_notex}).
 In particular, suppose that $\Lambda_1=|\varphi\>\<\varphi|$, then  $\pi(M)$ is
 extremal if and only if  $\ptr_{\Ka}(|\varphi\>\<\varphi|)=\sigma$. Indeed, 
this means  $\Lambda_2\in \Pe_{K_c}(\Ka\otimes\Ha)$ and 
$\Lambda_1\in  \Pe_{K_c}(\Ka\otimes\Ha)$ by Lemma \ref{lemma:gjrho} (i).
Since $\sigma$ has full rank, $\varphi$ is not a product vector and hence $M$ is
 an extremal 1-tester, by Example \ref{ex:extremal_qubit}. The converse is clear.
In particular, if $\sigma=1/2 I_{\Ha}$, then $M$ is extremal if and only if
 the vector $\varphi$ is maximally entangled.

On the other hand, if $\varphi$ is not a product vector but also not maximally entangled,
 then $M$ is an extremal qubit 1-tester such that $\pi(M)$ is not extremal.

\end{ex}

\section*{Acknowledgement}
I wish to thank the referee of the first version of this paper for very useful comments that greatly improved the paper.
This research was supported by the grants VEGA 2/0059/12, meta-QUTE ITMS   26240120022, Center of Excelence SAS - Quantum Technologies and by 
the Slovak Research and Development Agency
under the contract No. APVV-0178-11.

\end{document}